\documentclass[12pt,preprint]{aastex}

\slugcomment{\it Accepted for Publication in the Astrophysical Journal}
\shortauthors{Yan et al.}
\shorttitle{{\it Spitzer}/IRS Mid-Infrared Spectroscopy of $z \sim 2$ ULIRGs}

\def\deg{\ifmmode {^{\circ}}\else {$^\circ$}\fi}
\def\kms{\ifmmode {\rm\,km\,s^{-1}}\else
    ${\rm\,km\,s^{-1}}$\fi}
\def\ergcm2s{\ifmmode {\rm\,ergs\,cm^{-2}\,s^{-1}}\else
    ${\rm\,ergs\,cm^{-2}\,s^{-1}}$\fi}
\def\ergAcm2s{\ifmmode {\rm\,ergs\,cm^{-2}\,s^{-1}\,\AA^{-1}}\else
    ${\rm\,ergs\,cm^{-2}\,s^{-1}\,\AA^{-1}}$\fi}
\def\ergs{\ifmmode {\rm\,ergs\,s^{-1}}\else
    ${\rm\,ergs\,s^{-1}}$\fi}
\def\kmsMpc{\ifmmode {\rm\,km\,s^{-1}\,Mpc^{-1}}\else
    ${\rm\,km\,s^{-1}\,Mpc^{-1}}$\fi}

\def\spose#1{\hbox to 0pt{#1\hss}}
\def\simlt{\mathrel{\spose{\lower 3pt\hbox{$\mathchar"218$}}
     \raise 2.0pt\hbox{$\mathchar"13C$}}}
\def\simgt{\mathrel{\spose{\lower 3pt\hbox{$\mathchar"218$}}
     \raise 2.0pt\hbox{$\mathchar"13E$}}}

\def\plotfiddle#1#2#3#4#5#6#7{\centering \leavevmode
\vbox to#2{\rule{0pt}{#2}}
\includegraphics{#1}}

\begin{document}

\title{{\it Spitzer} Detection of PAH and Silicate Dust Features 
in the Mid-Infrared Spectra of $z\sim2$ Ultraluminous Infrared Galaxies}

\author{Lin Yan}
\author{R. Chary, L. Armus, H. Teplitz, G. Helou, D. Frayer, D. Fadda, J. Surace, P. Choi}
\affil{{\it Spitzer} Science Center, California Institute
of Technology, MS 220-6, Pasadena, CA 91125}

\email{lyan@ipac.caltech.edu}

\begin{abstract}

We report the initial results from a Spitzer GO-1 program to obtain
low resolution, mid-infrared spectra of infrared luminous galaxies
at $z \sim 1 - 2$. This paper presents the spectra of eight
sources observed with the {\it Spitzer} InfraRed Spectrograph (IRS).
Of the eight spectra, six have mid-IR spectral features, 
either emission from Polycyclic Aromatic Hydrocarbon
(PAH) or silicate absorption. Based on these mid-IR features,
the inferred six redshifts are in the range of $1.8-2.6$.
The remaining two spectra detect only strong continua, thus do not 
yield redshift information. Strong, multiple PAH emission 
features are detected in two
sources, and weak PAH emission in another two. 
These data provide direct evidence that PAH molecules
are present and directly observable in ULIRGs at $z \sim 2$. 

The six sources with measured redshifts are
dusty, infrared luminous galaxies at $z \sim 2$ with estimated 
$L_{bol} \sim 10^{13}L_\odot$. Of the eight sources, 
two appear starburst dominated; two with only power law 
continua are probably type I QSOs; 
and the remaining four are likely composite systems containing
a buried AGN and a starburst component.
Since half of our sample are optically faint sources with 
$R \ge 25.5$~mag (Vega), our results demonstrate the 
potential of using mid-infrared 
spectroscopy, especially the Aromatic and silicate features
produced by dust grains to directly probe optically faint and
infrared luminous populations at high redshift. 

\end{abstract}


\section{Introduction}

Numerous studies over the past decade have found
that star-forming galaxies have undergone
strong evolution in their luminosity and number density since redshift of $1 - 2$,
and that luminous and ultraluminous infrared galaxies play a critical role in this
evolution. The peak of the far-infrared background detected by {\it COBE}
is at $\sim200\mu$m, with energy comparable to 
the optical/UV background \citep{puget96,fix98}.
This implies that $\sim$50\%\ of the integrated rest-frame optical/UV
emission in the universe has to be thermally reprocessed by dust and radiated
at the mid to far-infrared. Deep ISO 15$\mu$m number counts,
and more recently, {\it Spitzer} 24$\mu$m number counts,
suggest a large excess of mid-IR sources compared to
the predictions by nonevolving models \citep{marleau04,casey04,chary04,Gruppioni02}.
Particularly, the {\it Spitzer} 24$\mu$m number counts
imply that a significant population of sources with flux densities
on the order of 100$\mu$Jy are likely infrared 
luminous galaxies at $z \sim 1 - 3$, previously undetected in 
the ISO data. Furthermore, the strong evolution of infrared luminous galaxies
at $z \simlt 1$ was directly characterized by the infrared luminosity functions
from studies of ISO 15$\mu$m sources with optical redshifts 
\citep{Elbaz99, Serjeant00,Franceschini02,Elbaz02},
In addition, optical spectroscopy of 
optically faint, radio-selected, S$_{850}>6$~mJy submillimeter
galaxies has found that these sources are at a median
redshift of 2.3 \citep{scott05,scott03}. Their inferred
volume density, $\sim1.3\times 10^{-5}{\rm Mpc^{-3}}$ at
$L_{IR} \ge 4\times 10^{12}L_\odot$,
is roughly two to three orders of
magnitude higher in comparison with the local density
of IRAS ultraluminous galaxies at the similar luminosity limit \citep{tom87}.

Although ISO and sub-mm observations have revealed 
many tantalizing facets of the dusty universe at $z \sim 1 - 3$, 
the successful launch of {\it Spitzer},
with its combined fast photometric
mapping and spectroscopic capabilities,
has made it possible to discover and characterize
large numbers of dusty, infrared
luminous galaxies at cosmologically interesting redshifts.
Of critical significance is that the mid-IR spectroscopic properties
of infrared luminous galaxies at $z > 0.5$ are virtually unknown.
The mid-IR spectra of starburst galaxies and most ULIRGs
are dominated by the emission and absorption features of dust
grains.  While the shape of the mid-IR continuum and its brightness
relative to the far-infrared constrain the amounts of hot and cold dust, 
the strong PAH emission features at 6.2, 7.7, 8.6,
11.3, and $12.7\mu$m and silicate
absorption (centered at $9.7$ and $18\mu$m) provide both an indication of
the type of source heating the dust (since PAH's are easily destroyed by
UV photons and x-rays from an AGN) and redshift estimates for sources that
are completely obscured at shorter wavelengths
\citep{genzel00,draine03,laurent00,tran01,dim99,voit92}.
With the sensitivity and wavelength coverage of the
InfraRed Spectrograph \citep[IRS;]{houck04} on {\it Spitzer},
it is now possible to obtain the mid-IR spectral diagnostics of dusty galaxies
out to $z \sim 3$. 

In this paper, we report initial results of a GO-1 program to obtain 
low resolution, mid-infrared 
spectra of a sample of 52 galaxies in the {\it Spitzer} First
Look Survey (FLS)\footnote{For
details of the FLS observation plan and the data release, see
http://ssc.spitzer.caltech.edu/fls.} over an area of 3.7deg$^{2}$. 
Our targets were potential high redshift candidates, selected based
on the color criteria described in detail in \S 2.
Eight sources in our target sample were observed
between August 27 to September 6, 2004, while the remainder of the full 
sample has been scheduled for 2005. In this paper, we report
the first results from the analyses of these eight targets. 
These eight sources have 24$\mu$m fluxes and mid-IR colors
representative of the full sample, and provide an initial look at
the data we expect to obtain for the complete sample.
Since very few mid-IR spectra of galaxies with $z > 1$ have been
published, and the lifetime of Spitzer is short, these initial results
are valuable for planning future IRS observations of high-redshift,
luminous infrared galaxies. The paper is organized as follows. \S 2 describes the 
our target selection and the IRS observation design in detail; \S 3
discusses the imaging reduction and 1D spectral extraction 
procedures we used in this dataset; \S 4 presents the results and we 
discuss the implications of our findings in \S 5.
Throughout the paper, we adopted the cosmology of $\Omega_m = 0.27,
\Omega_\Lambda = 0.73$ and $H_0 = 71$\kmsMpc.

\section{Target Selection and Observations}

Our {\it Spitzer} GO-1 program was designed to 
obtain mid-infrared spectra of a sample of 
24$\mu$m sources, selected to be starburst candidates
at $z \simgt 1 - 2$ based on their mid-IR colors. 
The {\it Spitzer} data used for our target selection
are the 24 and 8$\mu$m catalogs
from the FLS main survey region \citep{fadda05, marleau04, lacy05}.
In order to obtain good quality IRS spectra in reasonably 
short integration times, we required all
spectroscopic targets to be brighter than 0.9mJy at 24$\mu$m. 
We applied additional color constraints in order to select
potential starburst galaxies at $z > 1$. This is based on $24/8$ and $24/0.7$~micron colors.
Here we define $R(24,8) \equiv \log_{10}(\nu f_{\nu}(24\mu m)/\nu f_{\nu}(8\mu m))$,
and $R(24,0.7) \equiv \log_{10}(\nu f_{\nu}(24\mu m)/\nu f_{\nu}(0.7\mu m)$).
The criteria used to select the IRS targets are $R(24,8) \ge 0.5$
and $R(24,0.7) \ge 1.0$. These color cuts
are determined by comparing the observed 24/8 and 24/0.7~$\mu$m color distributions
of {\it all} 24$\mu$m sources over the FLS region with the 
computed color tracks as a function of redshift using a set of known
mid-IR spectral templates. The detailed discussion
on the mid-IR color distributions in the FLS has been published
in \citet{yan04}. The $R(24,0.7) \ge 1.0$ condition is a crude redshift selection,
and the $R(24,8) \ge 0.5$ cutoff selects sources with 
very steep, red continua, as seen in many local starbursts and ULIRGs 
\citep{armus04, genzel98,lutz98,dim99}. In \S 5, we will discuss how well our selection 
criteria work with respect to our observations. 

Over the 3.7deg$^2$ region, a total of 59 sources meet the 24$\mu$m flux
and the color cuts. Due to the limited number of hours available
to our program, we targeted 52, chosen randomly from
the total 59 sources. For all of the targets, we obtain low resolution 
(${\lambda \over \Delta \lambda} = 64 - 128$) spectra in 
Long-low module of the IRS (LL, 14 -- 40$\mu$m).
For a few sources with the IRAC 8$\mu$m fluxes
greater than $150\mu$Jy, we also obtained 1st order, Short-low
spectra (SL, 7.5--14$\mu$m). Table 1 lists the positions, 
the broad band 24$\mu$m, 8$\mu$m and R-band fluxes,
and the exposure times for these eight sources. 
In all of the observations, we used nearby bright stars as peak-up
to accurately center the slit on the science targets. 
The IRS slit width is $3.7^{''}$ for the SL and $10.7^{''}$ for the LL. 
Our targets are all unresolved, point sources with the full-width-half-maximum
(FWHM) of $\sim 2^{''}$ and $\sim 6^{''}$ in the 8 and 24$\mu$m images respectively.
Thus, in our IRS observations there should be
neglible light loss due to the finite IRS slit width. 

\section{Data Reduction and Analyses}

\subsection{Imaging Processing}

The raw spectral data are first processed by the IRS pipeline
(S11.0.2 version) at the Spitzer Science Center (SSC). 
The processing steps taken by the IRS pipeline include
ramp fitting, dark sky subtraction, droop correction, 
linearity correction, flat fielding, and wavelength and flux
calibration. For our SL data, stray light correction is not
necessary because the peak-up images do not have any sources bright enough to have
significant stray light contaminations in the SL spectra.
We therefore used the uncorrected output from the SSC pipeline.

Starting with the 2-dimensional (2D) Basic Calibrated Data (BCD)
produced by the IRS pipeline, we performed additional 
processing, including background subtraction and making 
new bad pixel mask for the coadded image at each
nod position. The background subtraction was tested with
three different methods: (1) The subtraction of a ``supersky",
constructed by median combining all of the available 2D images, excluding
the images at the same order and the same nod position.
This supersky is scaled
and then subtracted from the individual exposure before summing
up all exposures at each nod position,
(2) By differencing the exposures at two nod positions along the slit,
then summing up all the exposures at the same nod position, (3) 
The subtraction of a supersky constructed only
from the exposures within a single AOR.  The supersky methods (1 \&\ 3)
improve the background signal-to-noise ratio by $\sim$10-20\%.
However, for some fields, when the mean background levels
are significantly different from the supersky, the number of ``warm pixels''
can be higher than using the difference of the nod positions.
In these cases, we simply take the difference of the two nod positions.
In all cases, we chose the subtraction method that gives 
the best signal-to-noise ratio as well as
the fewest number of residual bad pixels.

After the background subtraction, we compare the 2D spectra of the same
target at the two nod positions and remove any bad pixels within
the spectral extraction aperture by visual examination. Usually, there are
less than 10 pixels which need to be masked manually in addition to the
bad pixel masks. These bad pixels are not considered in the subsequent, 1D
extracted spectra, and do not contribute to the final average of the 
spectra from the two nod positions.

\subsection{Spectral Extraction}

As described in \S 2, spectra were taken in two nod positions for each IRS slit. 
One-dimensional (1D) spectra were extracted, then averaged to produce the final
1D spectrum for each module. 
The spectral extraction is done using the SSC software SPICE v1.3\footnote{
see the ssc webpage http://ssc.spitzer.caltech.edu/postbcd/spice.html 
for the description of the software and the public release status.}.
SPICE is a Java based tool which allows the user to interactively
display and extract IRS spectra. The spectral centroiding and extraction aperture 
can be adjusted to maximize the S/N
ratio. For our data, we used an extraction aperture, whose width is 
1 or 2 pixels narrower than the default PSF size in SPICE, and the shape of 
the extraction aperture simply follows the PSF size as a function
of wavelength. This method minimizes the uncertainties due to the background, 
and maximizes the S/N ratios for faint spectra.
We used the simple method of summing up the light within the 
extraction aperture and did not use optimal extraction technique for our data.
The version of all of the calibration files used by SPICE is 
consistent with the IRS pipeline
version S11.0.2. The flux scales of the final, averaged SL and LL spectra 
were set by using the IRAC 8$\mu$m and MIPS 24$\mu$m broad band
fluxes. The accuracy of our flux calibration
is thus primarily determined by the IRAC 8$\mu$m and MIPS 24$\mu$m calibration errors, 
roughly (5-10)\% (see the {\it Spitzer} Observers Manual;
also Reach et al. in prep).

\section{Results}

\subsection{Spectra and Redshift Measurements}

Figure 1a,b,c,d show the 1D spectra of the eight sources 
observed by the IRS. In Figure 1, the color coded lines represent the 1D spectra
in the original resolution, with the red segment for the IRS LL 1st order, the blue 
for the LL 2nd order, and the green for the SL 1st order.
To aid the identification of PAH emission features,
we smoothed our observed 1D spectra by roughly
5-7~pixels. The smoothed spectra are shown by the black
lines, and are shifted by an arbitary amount in the Y-axis for
viewing clarity. The smoothing kernel is a boxcar, and roughly
$5-7$~pixels wide. The size of the smoothing kernel is chosen 
so that at the rest-frame wavelength it is similar to 
the the rest-frame FWHM of the 7.7$\mu$m PAH emission
in UGC~5101 \citep{armus04}.
Strong, multiple PAH emission features (6.2,7.7,8.6 and 11.2$\mu$m) are detected in
two sources, IRS9 and IRS2, and weak PAH emission (7.7$\mu$m and 6.2$\mu$m or 8.6$\mu$m) 
in additional two (IRS6 and IRS11). A broad absorption trough, roughly
centered at the observed wavelength of 27-36$\mu$m, is also detected in 
six out of eight spectra. We associate this broad feature
with redshifted silicate absorption at a central rest-frame wavelength of 9.7$\mu$m.
Based on these emission and absorption features, the inferred redshifts
are in the range of $1.8<z<2.6$ for these six sources.  
The remaining two spectra (Figure 1d) have well detected continua, but no other
identifiable spectral features. The redshifts for these two sources are unknown.

For the six sources with either PAH emission and/or silicate absorption
at the rest-frame 9.7$\mu$m, we measure their redshifts using the following procedures. 
Two of our sources have only broad silicate absorption troughs, 
thus, their redshift measurements have large errors, on the order of 0.1-0.2. 
In the remaining four sources, we have multiple spectral features, 
both PAH emission and silicate 
absorption. PAH features are generally sharper than silicate absorption, thus
give more accurate measurements of redshifts. 
In each of these cases, the final redshift is determined by
the combined $z$ measurements from the multiple features.
Specifically, for each PAH feature, the redshift $(1+z)_{i}$ is derived 
as $\lambda_{obs}(i)/\lambda_{rest}(i)$.
Here $i$ indicates one of the PAH features, and its rest-frame wavelength is measured 
from the spectra of local starburst galaxies taken with the 
IRS in the same slit and resolution. 
The local reference spectra were NGC7714 and UGC5101 
(Brandl et al. 2004; Armus et al. 2004).
The final redshift is the average value of $z_{i}$, and the 
error is computed as $\sigma = {\sqrt{\sum_i (z_i - <z>)^{2}} \over \sqrt{n-1}}$. 
The redshift discrepencies between different PAH features reflect the contributions
from the wavelength measurement error in determining
the centroid of PAH feature and the systematic errors in the IRS wavelength calibration.
Sharper PAH emission features will have smaller measurement errors in the wavelength
centroids. 
The redshifts for the six sources are in the range of 1.8-2.6.
The redshift errors for IRS2 and IRS9 are smaller, since both
have PAH emission lines at 6.2, 7.7, 8.6 and 11.2$\mu$m.
IRS9 has four strong PAH features. Its 6.2 and 11.2$\mu$m features are
narrow and used for the final redshift determination. The spectrum for IRS11 is peculiar. 
Its absorption line at 9.7$\mu$m is saturated with a flattened bottom. 
This is reflected in the
large uncertainty, 0.2, for its redshift.

An alternative way of determining the redshift is to 
fit the observed spectra using a set of mid-infrared spectral
templates of local galaxies with prominent PAH and silicate features 
(template cross-correlation method).
We found that the redshifts measured from these two methods
are consistent. The limitation in the template cross-correlation method
is that it has to assume some templates and it is difficult to determine 
what local template spectra are representative of high redshift spectra. 
For this paper, we are limited to a few, high quality spectra which have
been published.
We have used the interpolated model templates from Chary \&\ Elbaz
(2001), as well as the published IRS spectra of UGC~5101, Mrk1014, IRAS~F00183-7111
and NGC7714 \citep{armus04,spoon04,brandl04}. The strongest signal for determining 
redshift using the template fitting method comes from the silicate 
absorption feature. For this paper, we 
choose to use redshifts measured from PAH and silicate features. 
Table 2 presents the redshifts and the errors
for the six targets. 

\subsection{PAH Emission and Silicate Absorption Features}

Multiple, strong PAH emission features are detected in 
two sources and weak PAH emission in additional two.
Table 2 lists the PAH flux and its $S/N$ for each feature. 
The flux is measured by fitting a gaussian profile with continuum subtracted, and
a linear continuum computed over roughly a wavelength range of $\lambda_0 \pm 1.5\mu m$
($\lambda_0$ being the fitted PAH feature).
Given our low S/N spectra, we have adopted this simple approach to
measure the continuum and the line flux.
The $1\sigma$ noise for each PAH feature is computed as
$1\sigma = rms \times \sqrt{n} \times \Delta \lambda$,
where $rms$ is the noise per pixel measured from the continuum near the emission line,
$n$ is the FWHM of the emission line in pixel, and $\Delta \lambda$ is the size of
the pixel in $\mu$m/pixel. The significance of the detection (S/N)
is computed as $S/N = flux(PAH)/1\sigma$. 
The $rms$ is measured from the original spectrum and the feature
profile fitting is done with the smoothed spectrum.
As shown in Table 2, the $S/N$ ratio for each individual PAH feature is geater than 3
for IRS2 and IRS9, and in the range of $1.5 - 2.5$ for IRS6 and IRS9.
Although individual $S/N$ ratio seems to be small, the final redshift
measurement is much more reliable when the identified, multiple features
are consistent. 

One pecularity about the IRS9 spectrum is its unusually strong 11.2$\mu$m
line, in comparison with UGC5101, F00183-7111 and NGC7714
\citep{brandl04, armus04, spoon04}.  The relative
line luminosity between 6.2 and 11.2$\mu$m is 0.75 for IRS9.  One study in
local HII regions \citep{vermeij02} seems to suggest that this ratio less
than 1 are mostly in low metallicity objects. In addition, the theoretical
models of PAH emission \citep{draine01} show that luminosity ratio of
11.2-to-7.7$\mu$m features is closely linked to ionization state of PAH,
and the high ratio is indicative of presence of a large amount of neutral
PAH, with ISM similar to photo-disassociate-region (PDR).

The strength of silicate absorption is an excellent 
indicator of dust extinction at the mid-infrared. 
Precise measurement of the absorption strength sometimes can be 
difficult when the feature is not very strong and the continuum
is affected by strong PAH
emission at the both side of the absorption trough.
Of the six targets with redshifts, 
at least four have strong silicate absorption. In the other two sources, 
IRS9 and IRS11, the absorption due to silicate material is probably present, but
the estimate of the depth of the absorption is largely 
uncertain. This is mainly due to the poor determination
of the continuum level using the low $S/N$ spectra covering a fairly narrow 
wavelength region.

Although comprehensive modeling of our mid-IR spectra is 
beyond the scope of this paper, we can use the following
methods to provide a simple estimate of the dust obscuration in our sources. 
The depth of the absorption feature
can be used as an indication of the obscuration. The opacity at the bottom of 
the silicate absorption trough is estimated
as $\tau_{9.7\mu m} = \ln(S_c/S_{min})$, where $S_c$ and $S_{min}$
are the flux densities at the center of the trough without and with absorption
respectively. The continuum is measured using clean continuum location at rest-frame
7 and 14$\mu$m. $S_{min}$ is taken as the minimum of the flux density at
the bottom of the absorption trough. Table 2 lists $\tau_{9.7\mu m}$ values for the
six sources we detect the silicate absorption. 
The amount of dust obscuration is highly uncertain in IRS11.
Its spectrum shows that the absorption minimum is flattening out to zero, 
suggesting a large amount of dust along the line-of-sight.
However, the level continuum for this source could also be low, thus
the inferred absorption strength would be small. It requires a better
$S/N$ ratio spectrum to draw more definitive conclusion on its dust obscuration 
for this object. From $\tau_{9.7\mu m}$ listed in Table 2, the 
implied extinction at the visible wavelength is large, if we assume
$A_V/\tau_{9.7\mu} = 18.5 \pm 2.0$ \citep{draine03}. The estimated
lower limits on $A_V$ is $\sim 14-56$~magnitudes, which is comparable
to that of local ULIRGs.

Another method would be to use the absorption
coefficients computed by Li \&\ Draine (Li \&\ Draine 2001), based
on a simple cold screen model. The computed spectra, 
$I_\nu / I_\nu(cont) = e^{-N*C_{ext}}$, are used to fit
the observed spectra between in the rest-frame $6-12\mu$m.
The estimated column density is in the range of $(2-10)\times 10^{22}$cm$^{-2}$.
We should point out that the uncertainties in the quantative analyses of silicate
absorption feature and the dust extinction are mostly from
determination of continuum and unknown geometric distribution of dust. 
While both estimates of the extinction towards the line of sight
to the nuclei are valid, major uncertainties remain concerning
the distribution of the dust, the dust-to-gas ratio, the metallicity
and the proper level of the continuum.  Therefore, the derived values
should be treated as rather uncertain estimates of the lower limit of
extinction along the line of sight to the nuclei of these high-redshift
galaxies.

\subsection{Bolometric luminosity}

While our IRS spectra covering the rest-frame 5-15$\mu$m do provide constraints
on the slopes of the mid-IR SEDs, the far-infrared emission,
which dominates the bolometric luminosity, remains unconstrained
for most of our sample, except IRS9. For IRS9, we have 
a MIPS $10\sigma$ detection at 70$\mu$m with the flux of 42mJy.
At 160$\mu$m, we have only a $3\sigma$ upper limit of 150mJy. 
IRS9 is also detected at 1.2mm with 6$\sigma$ of 2.5mJy with MAMBO \citep{lutz05}.
Combining all of the data together, we estimate that $L_{IR}$ for 
IRS9 is $1.8\times10^{13}L_\odot$. The uncertainty for IRS9 is probably witin a
factor of 2-3. For the remaining five sources with redshifts, the crude 
estimates of their bolometric luminosities were done
using the template fitting method. As templates, we used the IRS spectra of
UGC5101, Mrk1014, IRAS~F00183-7111 and NGC7714 \citep{brandl04, armus04,
spoon04}.

The derived bolometric luminosities are in the range between $(0.5 -
5)\times 10^{13}L_\odot$.  These luminosities are likely uncertain to
factors of $5-10$ because the selection of the best-fit, rest-frame
mid-infrared spectrum from a limited number of templates does not ensure
an accurate estimate of the bolometric luminosity, especially the cold
dust component.  However, since the templates include pure starbursts,
composite sources, and pure AGN, they likely provide a fair representation
in bolometric luminosity, energetic type and degree of dust obscuration,
of the sources in our sample.  IRAS~F00183-7111 is very luminous, highly
obscured source, UGC 5101 has a powerful starburst and a buried AGN, Mrk
1014 is a type-1 dusty QSO, and NGC 7714 is a bright, nearby starburst
(not a ULIRG). Both Mrk 1014 and NGC 7714 have very little extinction
along the line of sight to their nuclei (as evidenced by the lack of
silicate absorption in their IRS spectra).  With these uncertainties in
mind, our derived luminosities suggest that the observed galaxies are as
luminous, and likely much more luminous, than typical low-redshift
ULIRGs.  More importantly, our sources with PAH emission features are a
factor of (5-10) more luminous than starburst dominated ULIRGs in the
local Universe.

\section{Discussions}

\subsection{The Nature of our sources}

With redshift information, we compute PAH luminosities and estimate
their contributions to the bolometric infrared luminosities. 
Table 3 lists the PAH-to-bolometric luminosity ratios and the 
rest-frame equivalent widths, including 
the data for the local benchmarks.
The $L_{7.7\mu m}/L_{IR}$ is $10^{-3}$ and $2.2\times 10^{-3}$ for 
IRS2 and IRS9 respectively. These values are similar to UGC5101, a highly obscured and
starburst dominant system. And they are much higher than those of Mrk1014
(type-I, IR loud QSO) and F00183-7111 (a highly obscured composite ULIRG).
The rest-frame equivalent widths at 6.2 and 7.7$\mu$m
are comparable to those of UGC5101 in IRS9.
In IRS2 the equivalent widths are smaller than in UGC 5101 bu
about a factor of three. The IRS spectra of IRS9 and to a lesser extent,
IRS2, both indicate a dominant contribution to the bolometric luminosity
from a dusty starburst (as in UGC 5101).  In IRS2 a buried AGN might be
contributing at IRS wavelengths, decreasing the equivalent width of the
PAH features by increasing the continuum from hot dust.  The
starburst-like nature of IRS9 is further supported by its detections at
70$\mu$m with {\it Spitzer} and at 1.2mm with MAMBO on IRAM 30~meter
telescope.

For comparison with published
ISO results on the PAH emission strength in ULIRGs, we measured 
the 7.7$\mu$m line-to-continuum ratio $(l/c)_{7.7\mu m}$
for IRS2 and IRS9. With two anchor points at the rest-frame 7 and 14$\mu$m,
the ratios are 1 and 1.3 for IRS9 and IRS2 respectively. An
$(l/c)_{7.7\mu m} \ge 1.0$ is shown by various ISO studies to imply
a dominant starburst contribution to the mid-infrared flux
\citep{genzel98,dim99}. Since both IRS9 and IRS2 meet this criterion, it
is likely that they are starburst-like ULIRGs.

The remaining four of the six sources with redshifts
have silicate absorption, but weak or no PAH emission.
In IRS6 and IRS11, the $S/N$ of individual PAH feature
is only at the level of $1.5-2.5$. 
IRS1 and IRS8 have no detectable PAH emission.
If we take the PAH fluxes or $1\sigma$ upper limits, and bolometric luminosities 
listed in Table 2, the implied 7.7$\mu$m line-to-bolometric luminosity ratios
are (2, 10, 8 and 6)$\times 10^{-4}$ for IRS1, IRS6, IRS8 and IRS11 respectively.
These ratios should be taken as upper limits. Compared with 
those of the local galaxies in Table 3, these numbers suggest
that IRS1 is probably a system mostly dominanted by an obscured AGN, whereas
the remaining three sources, IRS6, IRS8 and IRS11, are dusty, composite systems
with both AGN and starburst contributions to their infrared luminosities.

Finally, IRS4 and IRS10 have only continua without any 
identifiable spectral features. Their redshifts can not
be measured from the IRS spectra. Their observed IRS spectra
from 8-38$\mu$m can exclude the possibilities that these two objects
have {\it strong} PAH emission lines at both low and high redshifts.
The observed power law continua suggest that 
they are probably type I QSOs, with very little PAH emission at
either low or high redshifts. 

\subsection{Sample Selection and The Nature of the Sources}

Although our statistics are poor at this stage, we can
assess the effectiveness at which our color criteria
select for high-redshift, dusty galaxies.
Primarily, we select sources with very red 24-to-8 and 24-to-0.7 micron colors.
These red color cuts are chosen to target dusty, starburst galaxies at $z \ge 1$.
The IRS GTO team has recently published another sample of $z \sim 2$ ULIRGs
with the Spitzer mid-IR spectra (Houck et al. 2005).
One of the differences between our sample and the Houck et al. sample 
is in the initial target selection. We utilized IRAC 8$\mu$m photometry and
applied uniform flux ratio cuts in two colors, whereas the Houck et al. sample
targeted only optically faint, 24$\mu$m sources. 
Particularly, our red 24-to-8~$\mu$m color criterion selects sources with
red, mid-IR continua. Of the eight targets, $75\pm31$\% have redshifts of $z > 1$.
Our selection clearly works well in targeting high redshift systems, and results in
a higher fraction of sources with measurable redshifts based on the mid-IR
spectra than that of the Houck et al. sample.
In our small sample, at least four and possibly six
sources show strong silicate absorption, indicating large columns of 
cold dust along the line-of-sight to the nuclei. 
While a red, steeply rising mid-infrared continuum suggests the presence of warm
dust, the detected silicate absorption indicates also the cold dust. 
Our result may suggest a broad correlation between red mid-IR continua
and silicate absorption. Further studies with larger samples 
are clearly needed. Finally, our data detected, for the first time, strong
multiple PAH emission features in starbursts at $z \sim 2$, and our sample 
seems to have a higher fraction of systems with PAH emission than the Houck et al.
sample. This could be due to the differences in the sample selection functions. 

Although our sample selection was designed to target specifically starburst galaxies
with strong PAH emission, the fraction of such sources in our sample 
is rather small, $\sim25$\%\ from this small subsample. Most of our targets seem to have either
undetected, or weak, PAH emission compared to low-redshift ULIRGs.
This is perhaps not surprising given the luminosity of our targets,
and the fact that at low redshift, there appears to be a trend
toward more AGN-like mid-IR spectra at luminosities above
$5x10^{12}L_\odot$ (Tran et al. 2001; Charmandaris et al. 2005).
A similar trend is seen in $z \sim 2$ sources with the IRS selected
via different color criteria (Houck et al. 2005).
We should emphasize that although these extremely luminous
galaxies have AGN-like mid-IR spectra with weak PAH emission, 
they could still have some amount of starburst components. 
Local examples of this type of sources are Mrk231 and IRAS F00183-7111
\citep{spoon04}. Although Mrk231 has
weak PAH emission, its $L_{x}/L_{IR}$ ratio is much smaller than
that of a typical AGN-dominant or QSO \citep{bravit04}, indicating that starburst
contributes to the bolometric infrared luminosity. 
Our results suggest that we need to reach fainter 
apparent mid-infrared flux levels to uncover larger
populations of starburst-dominated sources at these redshifts
with the IRS.


For comparison, the comoving number
density of sub-mm detected galaxies at $<z>=2.2$ is roughly
$6\times 10^{-6}{\rm Mpc^{-3}}$ for $L_{ir} \ge 4\times 10^{12}L_\odot$
\citep{scott04}, and the the rest-frame UV color selected galaxies
have co-moving space density of $2\times 10^{-3}{\rm Mpc^{-3}}$
for the Balmer break selected galaxies at $<z>=1.77$ and $<z>=2.32$
\citep{adelberger04a}. Although our current sample is very small, 
it is illustrative to compute the co-moving number density of infrared luminous galaxies
at $z \sim 2$. Of all of the 24$\mu$m sources detected over $3.7$deg$^{2}$,
59 sources meet our sample selection criteria. If we assume $75\pm31$\%\ of these 59
sources being at $z=2.0\pm0.3$, the comoving number density is 
$n \sim 2 \times 10^{-6} {\rm Mpc^{-3}}$ for $z = 2$ and $L_{IR} \ge 5\times10^{12}L_\odot$.
Given the large uncertainties, our estimate is probably
consistent with that of sub-mm sources
and is roughly 1\%\ of the UV selected $z\sim2$ galaxies.

\section{Summary}

This paper presents the mid-IR spectra of eight
sources from our {\it Spitzer} GO-1 program, designed to investigate
the spectroscopic properties of high-redshift 
dusty, infrared luminous galaxies. 
We detect mid-IR spectral features, either emission lines from PAH or
silicate absorption at 9.7$\mu$m, among six out of eight
targets. The inferred redshifts, based only on the mid-IR spectra,
are at $1.8-2.6$. The remaining two sources have 
strong continuum but no identifiable spectral lines.
The main conclusions based on these eight mid-IR spectra are as follows.

\noindent(1) Our sample selection is effective in targeting
ULIRGs at high redshifts with PAH emission
and/or silicate absorption. 
The high 24$\mu$m to R-band flux density ratio criterion 
filters out low redshift galaxies, while the 
high 24/8$\mu$m ratio favors the selection of galaxies with red, steeply
rising spectra. The sample yields the 
redshift measurement efficiency of $75\pm31$\%.
(2) More than half of our sample (5/8) has
prominent silicate absorption lines at 9.7$\mu$m,
indicative of high dust obscuration in these $z\sim2$ sources.
(3) We detect multiple PAH emission features from two sources (2/8=25\%).
Compared to local ULIRGs, their line-to-bolometric luminosity
ratios and equivalent widths suggest that they are 
powered by extremely luminous starbursts.
These starbursts at $z \sim 2$ are a factor of (5-10) more 
luminous than the local starburst dominated ULIRGs. 
These data are the first direct evidence that aromatic features
are already a significant component of the dusty galaxy spectra
as early as z=2.
(4) The remaining four of the six sources with redshifts (4/8=60\%) 
have weak or no PAH emission,
and are likely composite systems having both a starburst and a buried
AGN.  Although the statistics thus far are small, our selection criteria
appear to be good at isolating $z \sim 2$ ULIRGs with, in most cases,
spectra most resembling AGN-dominated ULIRGs at low-z.  However, these
data also suggest that small PAH emitting dust grains are present and
directly observable in up to $20-40$\% of our ULIRGs at $z
=2$.  Since the IRS integration times we have used for these data are
relatively short, this bodes extremely well for the goal of directly
using {\it Spitzer} to spectroscopically classify 
high-redshift luminous infrared galaxies.

\acknowledgements
We are grateful to the IRS instrument team and the IRS instrument support team
at the Spitzer Science Center for their tremendous effort to make
this instrument such a great success.
We thank Tom Soifer for stimulating discussions on the nature of these luminous sources.
Support for this work was provided by NASA through award issues by JPL/Caltech.

\clearpage

\begin{figure}[!t]
\begin{center}
\plotfiddle{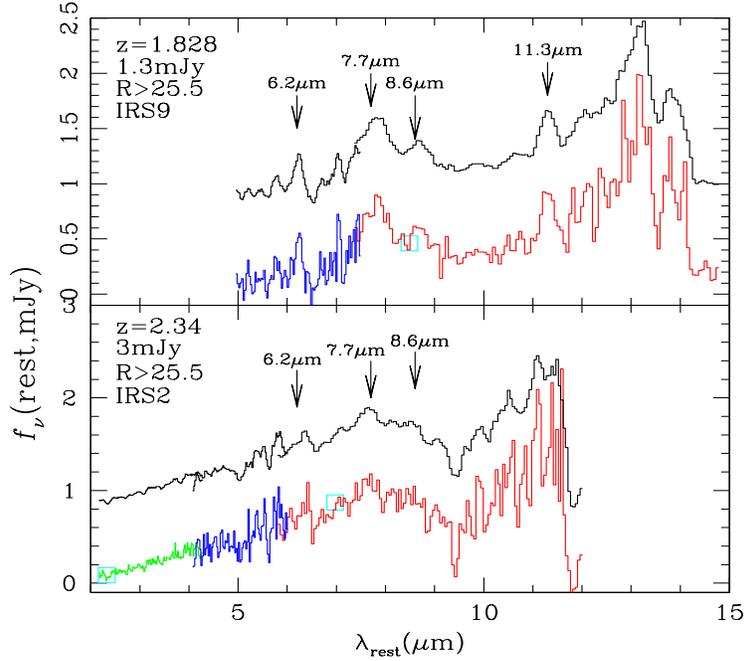}{2.8in}{0}{50}{45}{-150}{-95}
\end{center}
\caption{{\bf Panel a:} Here are two spectra presented in the rest-frame flux density
versus wavelength. The red, blue and green lines are the observed spectra in the original
resolution, whereas the black lines are the smoothed version, and scaled with an arbitary
constant in the Y-axis for viewing clarity. The red segment represents the spectrum from 
the IRS long-low 1st order, the blue for the 2nd order and the green is from the short-low
1st order. The open, cyan squares are the broad band rest-frame flux densities computed from
the observed, broad band 8$\mu$m from IRAC and 24$\mu$m fluxes from MIPS. }
\label{s:spectraa}
\end{figure}

\clearpage

\begin{figure}[!t]
\begin{center}
\plotfiddle{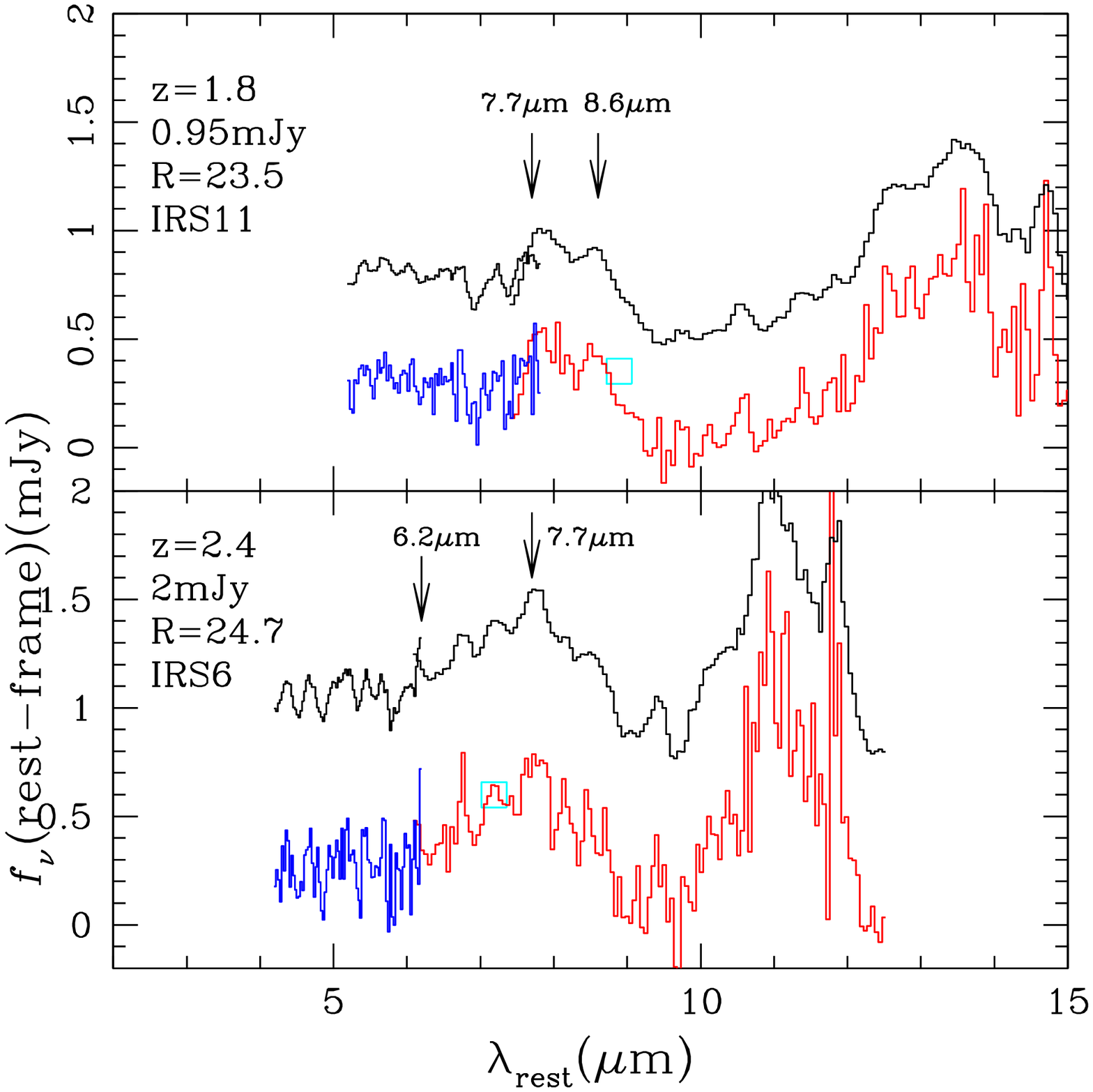}{3.8in}{0}{60}{50}{-150}{-90}
\end{center}
{{\bf Figure 1 - Panel b:} The caption is the same as in {\bf Panel a}.}
\label{s:spectrab}
\end{figure}

\clearpage

\begin{figure}[!t]
\begin{center}
\plotfiddle{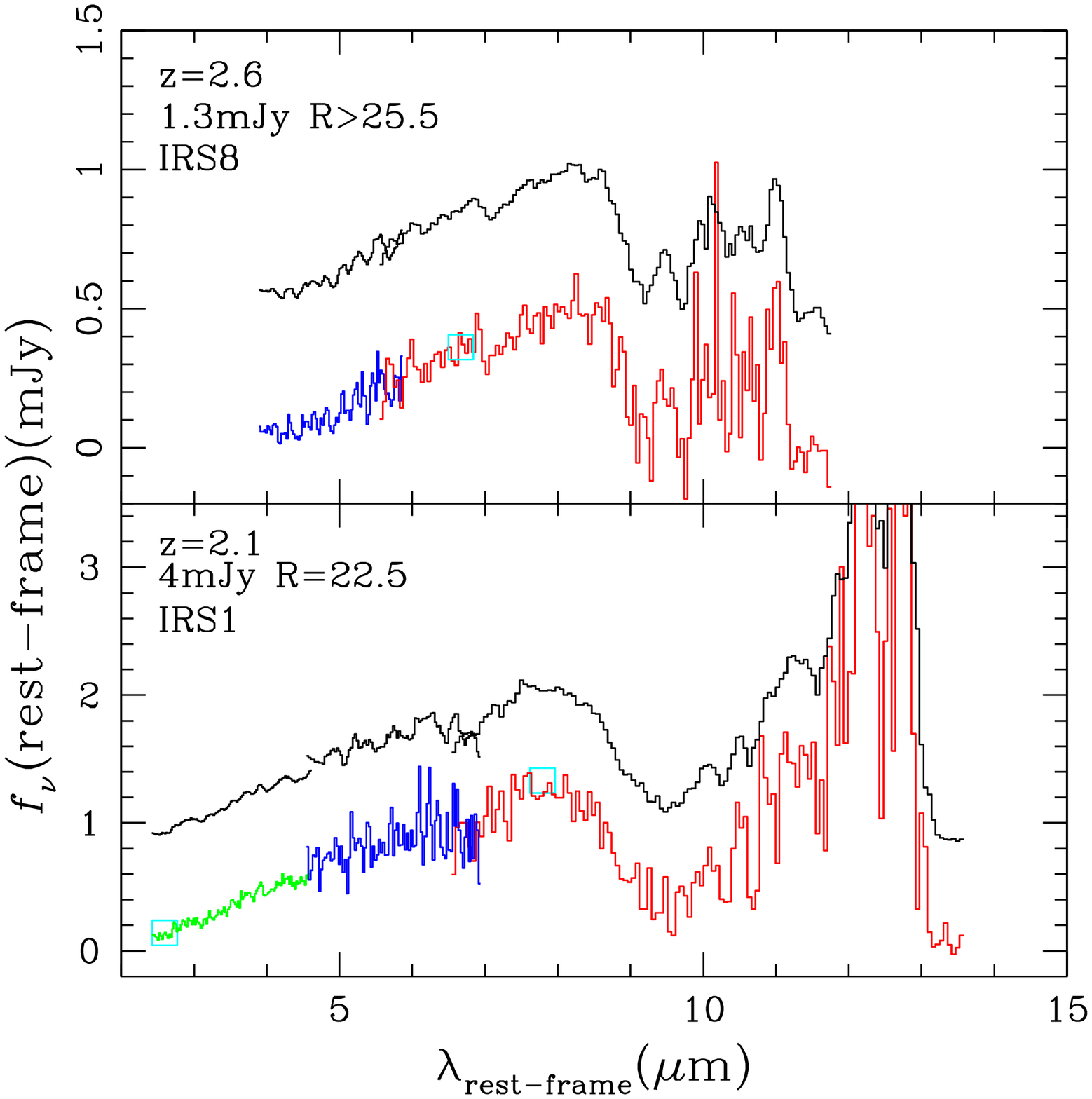}{3.8in}{0}{60}{50}{-150}{-90}
\end{center}
{{\bf Figure 1 - Panel c:} The caption is the same as in {\bf Panel a}.}
\label{s:spectrac}
\end{figure}

\clearpage

\begin{figure}[!t]
\begin{center}
\epsscale{.80}
\plotfiddle{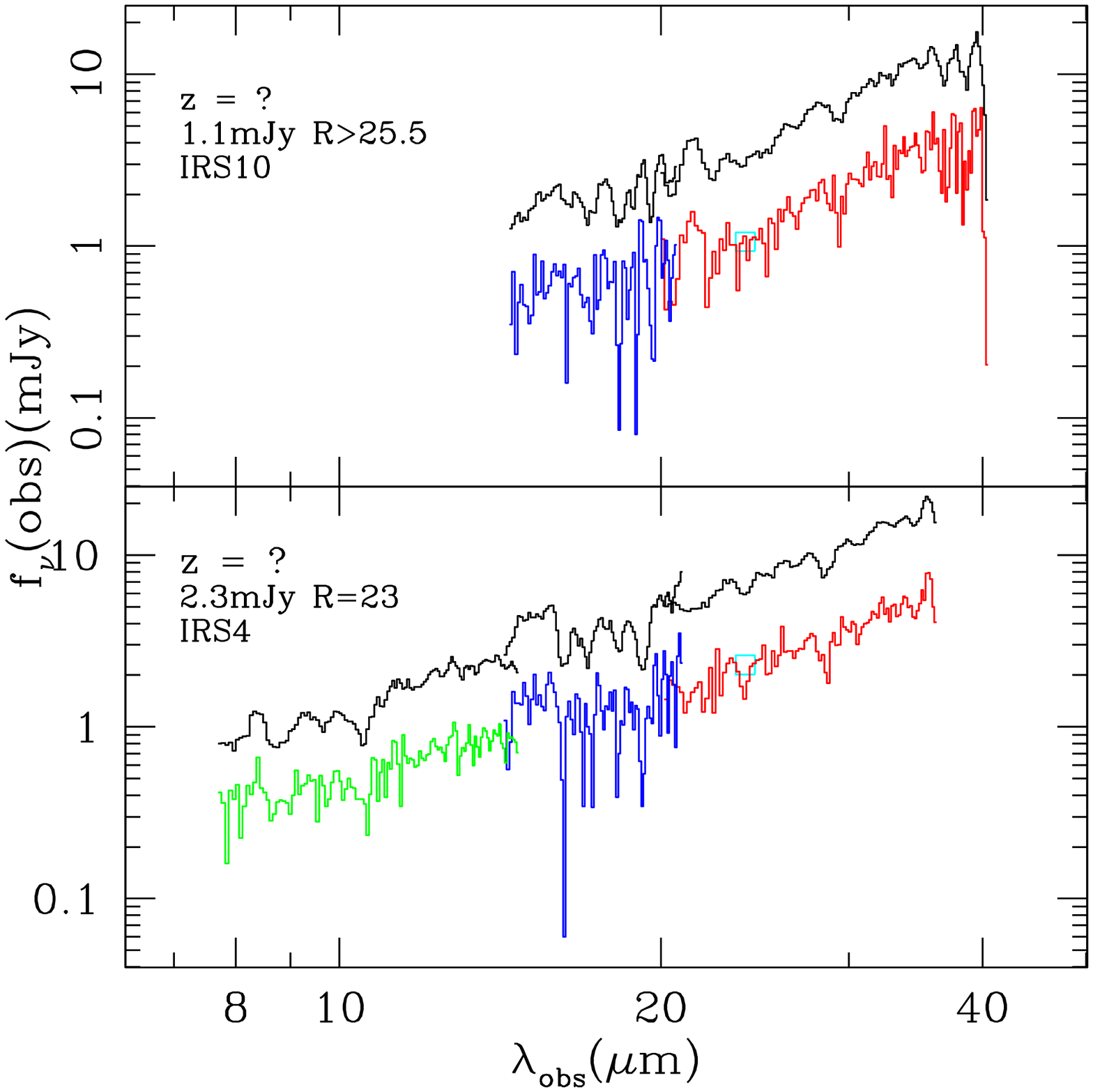}{2.8in}{0}{50}{45}{-150}{-90}
\end{center}
{{\bf Figure 1 - Panel d:} The caption is the same as in {\bf Panel a}.
These two spectra have no strong silicate absorption. IRS10 with weak
PAH emission is plotted in the rest-frame, and IRS4 with no identifiable
features is presented in the observed frame since the redshifts
are unknown.}
\label{s:spectrad}
\end{figure}

\clearpage


\clearpage
\begin{deluxetable}{ccccccccc}
\tabletypesize{\scriptsize}
\tablecaption{IRS Low Resolution Spectroscopy Observation Log}
\tablewidth{0pt}
\tablehead{
\colhead{ID} & 
\colhead{RA} &
\colhead{DEC} &
\colhead{$f(24\mu m)$} &
\colhead{$f(8\mu m)$} &
\colhead{$R$\tablenotemark{a}} &
\colhead{SL 1st} &
\colhead{LL 2nd} &
\colhead{LL 1st}  \\
   & (J2000) & (J2000) & mJy & $\mu$Jy & mag & seconds & seconds & seconds
}
\startdata
IRS1 & 17:18:44.378 & +59:20:0.53 & 4.1 & 428 & 22.5 & $1\times2\times240$ & $2\times2\times120$ & $3\times2\times120$ \\
IRS2 & 17:15:38.182 & +59:25:40.12 & 2.97 & 253 & $>$25.5 & $1\times2\times240$ & $2\times2\times120$ & $3\times2\times120$ \\
IRS4 & 17:12:15.439 & +58:52:27.88 & 2.3  & 261 & 23.0 & $1\times2\times240$ & $2\times2\times120$ & $3\times2\times120$ \\
IRS6 & 17:18:48.802 & +58:56:56.83 & 2.03 & 95.5 & 24.7 & \nodata     & $2\times2\times120$ & $3\times2\times120$ \\
IRS8 & 17:15:36.336 & +59:36:14.76 & 1.34 & $<$20 & $>$25.5 & \nodata & $6\times2\times120$ & $8\times2\times120$ \\
IRS9 & 17:13:50.001 & +58:56:56.83 & 1.28 & 93.5 & $>$25.5 & \nodata  & $6\times2\times120$ & $8\times2\times120$ \\
IRS10 & 17:21:24.581 & +59:20:29.51 & 1.06 & 28 & $>$25.5 & \nodata & $6\times2\times120$   & $8\times2\times120$ \\
IRS11 & 17:14:39.570 & +58:56:32.10 & 0.96 & 56 & 23.5 &  \nodata  & $6\times2\times120$   & $8\times2\times120$ \\
\enddata
\tablenotetext{a}{$R$ band magnitude is in Vega system. The 3$\sigma$
limit within an aperture of $3^{''}$ in diameter is 25.5mag.}
\end{deluxetable}

\clearpage

\begin{deluxetable}{cccccccc}
\tabletypesize{\scriptsize}
\tablecaption{Redshifts and spectral line measurements}
\tablewidth{0pt}
\tablehead{
\colhead{ID} &
\colhead{$z$} &
\colhead{$L_{ir}$} &
\colhead{$6.2\mu m$} &
\colhead{$7.7\mu m$} &
\colhead{$8.6\mu m$} &
\colhead{$11.2\mu m$} &
\colhead{$\tau_{9.7\mu m}$} }
\startdata
IRS1 & $2.1\pm0.1$  & $3.2\times10^{13}$  & \nodata & $<0.5 (1.0)$\tablenotemark{a} & \nodata & \nodata    & 1.77 \\ 
IRS2 & $2.34\pm0.06$  & $4.3\times10^{13}$ & $3.7 (3) $ & $3.2 (4.7)$ & $1.3 (3.8)$ & \nodata & 0.73 \\
IRS6 & $2.4\pm0.1$  & $1.3\times10^{13}$  & $1.3 (1.5)$\tablenotemark{b} & $1 (2.0)$ & \nodata & \nodata   & 1.43 \\
IRS8 & $2.6\pm0.2$   & $2\times10^{13}$   & \nodata & $<0.3 (1.0)$ & \nodata & \nodata   & 1.7 \\
IRS9 & $1.83\pm0.03$ & $1.8\times10^{13}$  & $4.0 (7.7)$ & $5.2 (6.6)\tablenotemark{b}$ & $2.7 (4.4)$ & $3.0 (13.1)$ & 1.2 \\
IRS11 & $1.8\pm0.15$  & $5.8\times10^{12}$ & \nodata  & $1.7 (2.5)$   & $1.3 (2)$ & \nodata & $>$3 \\
\enddata
\tablenotetext{a}{For PAH feature, we listed two numbers -- the first number is the observed line flux in unit
of $10^{-15}$ergs/s/cm$^2$, and the second number in the parenthesis 
is the signal-to-noise (S/N)
ratio of the flux. For the sources (IRS2 and IRS8) without detectable PAH features, we listed the 
$1\sigma$ flux limit for 7.7$\mu$m feature only. }
\tablenotetext{b}{The 6.2$\mu$m for IRS6 and the 7.7$\mu$m measurements for IRS9 are highly uncertain because
the lines are at the edges of the two orders in the long-low module.}
\end{deluxetable}

\clearpage
\begin{deluxetable}{cccccccccc}
\tabletypesize{\scriptsize}
\tablecaption{Comparison with local ULIRGs}
\tablewidth{0pt}
\tablehead{
\colhead{ID} &
\colhead{$z$} &
\colhead{$L_{IR}$\tablenotemark{a}} &
\colhead{$L_{6.2\mu} \over L_{IR}$ } &
\colhead{$L_{7.7\mu} \over L_{IR}$  } &
\colhead{$L_{11.2\mu} \over L_{IR}$  } &
\colhead{$EW^{rest}_{6.2\mu}\tablenotemark{b}$}  &
\colhead{$EW^{rest}_{7.7\mu}$}  &
\colhead{$EW^{rest}_{11.2\mu}$} & 
\colhead{$\tau_{9.7\mu}$\tablenotemark{c}} \\
\colhead{} &
\colhead{} &
\colhead{$L_\odot$} &
\colhead{ } &
\colhead{ } &
\colhead{ } &
\colhead{$\mu$m}  &
\colhead{$\mu$m}  &
\colhead{$\mu$m} &
\colhead{}}
\startdata
IRS2    & 2.34    & $4.3\times 10^{13}$  & 1.1e-3 & 1e-3 & \nodata & 0.08 & 0.13 & \nodata & 0.73 \\
IRS9    & 1.83   & $1.8\times10^{13}$  & 1.3e-3 & 2.2e-3\tablenotemark{d} & 1.2e-3 & 0.24 & 0.3 & 0.35 & 1.2 \\
\hline \\
F00183-7111 & 0.327 & $7\times10^{12}$  & $<$3.7e-4 & \nodata & 4.4e-4 & $<$0.011 & \nodata & 0.182\tablenotemark{d} & 5 \\
Mrk1014 & 0.1631 & $3.2\times10^{12}$   & 6.5e-5 & 1.5e-4  & 2.1e-5 & 0.048 & 0.022 & 0.017 & 0 \\
UGC5101 & 0.0394 & $8.7\times10^{11}$   & 1.9e-3 & 5.5e-3  & 1.0e-3 & 0.248 & 0.553 & 0.273 & 2 \\
NGC7714 & 0.0093 & $5.6\times10^{10}$   & 3.9e-3 & 7.9e-3  & 1.8e-3 & 0.50  & 0.70  & 0.17  & 0 \\
\enddata
\tablenotetext{a}{The total bolometric luminosity
$L_{IR}$ is for $3 - 1000\mu$m, and the unit is in $L_\odot$.}
\tablenotetext{b}{the rest-frame equivalent width is in micron.}
\tablenotetext{c}{Here $\tau_{9.7\mu}$ is computed from the minimum of the 9.7$\mu$m absorption, without
fully integrating over the line. This value should be the lower limit.}
\tablenotetext{d}{The data for the local galaxies 
in this table are from \citet{spoon04}, \citet{brandl04}, \citet{armus04}.
The rest-frame equivalent widths for F00183-7111 is from the private communication with H. Spoon (2004).}
\end{deluxetable}


\begin{thebibliography}{}

\bibitem[Adelberger et al.(2004)]{adelberger04a} Adelberger, K.~L., Steidel, C.~C.,
Pettini, M., Shapley, A.~E., Reddy, N.~A., Erb, D.~K.\ 2004, \apj, in press (astro-ph\/041065)

\bibitem[Alexander et al.(2003)]{alex03} Alexander, D.~M., et al.\ 2003, \aj, 125, 383

\bibitem[Armus et al.(2004)]{armus04}
Armus, L. et al. 2004, ApJS, 154, 178

\bibitem[Bohlin et al.(1978)]{bohlin78} Bohlin, R.~C., Savage,
B.~D., \& Drake, J.~F.\ 1978, \apj, 224, 132

\bibitem[Brandl et al.(2004)]{brandl04}
Brandl, B.R. et al. 2004, ApJS, 154, 188

\bibitem[Braito et al.(2004)]{bravit04} Braito, V., et al.\ 
2004, \aap, 420, 79 

\bibitem[Chary et al.(2004)]{chary04}
Chary, R. et al. 2004, ApJS, 154, 80

bibitem[Chary \&\ Elbaz (2001)]{chary01}
Chary, R. \&\ Elbaz, D. ApJ, 556 562

\bibitem[Chapman et al.(2005)]{scott05} Chapman, S.~C., Blain, 
A.~W., Smail, I., \& Ivison, R.~J.\ 2005, \apj, 622, 772 

\bibitem[Chapman et al.(2004)]{scott04} Chapman, S.~C., Smail,
I., Blain, A.~W., \& Ivison, R.~J.\ 2004, \apj, 614, 671

\bibitem[Chapman et al.(2003)]{scott03}
Chapman, S.;Blain,A.W.;Ivison,R.J.;Smail,I.R. 2003, Nature, 422, 695

\bibitem[Draine \& Li(2001)]{draine01} Draine, B.~T.~\& Li, A.\
2001, \apj, 551, 807

\bibitem[Draine(2003)]{draine03} Draine, B.~T.\ 2003, \araa, 41,
241

\bibitem[Elbaz et al.(1999)]{Elbaz99}
Elbaz et al. 1999, A\&A, 351, 37

\bibitem[Elbaz et al.(2002)]{Elbaz02}
Elbaz,D.;Cesarsky, C.J.;Chanial,P.;Aussel,H.;Franceschini,A.;Fadda,D;Chary,R.R. 2002, A\&A, 384, 848

\bibitem[Fadda et al.(2005)]{fadda05}
Fadda, D. et al. 2005, ApJ, in prep

\bibitem[Fixsen et al.(1998)]{fix98}
Fixsen,D.J.; Dwek,E.; Mather,J.C.;Bennett,C.J.;Shafer,R.A. 1998, ApJ, 508, 123

\bibitem[F{\" o}rster Schreiber et al.(2001)]{natasha01} F{\" 
o}rster Schreiber, N.~M., Genzel, R., Lutz, D., Kunze, D., \& Sternberg, 
A.\ 2001, \apj, 552, 544 

\bibitem[Franceschini et al.(2002)]{Franceschini02}
Franceschini et al. 2002, ApJ, 568, 470

\bibitem[Genzel et al.(1998)]{genzel98} Genzel, R., et al.\
1998, \apj, 498, 579

\bibitem[Genzel \& Cesarsky(2000)]{genzel00}
Genzel, R. \& Cesarsky, C.J. 2000, ARA\&A, 38, 761

\bibitem[Gruppioni et al.(2002)]{Gruppioni02}
Gruppioni et al. 2002, MNRAS, 335, 831

\bibitem[Houck et al.(2004)]{houck04} Houck, J.~R., et al.\
2004, \apjs, 154, 18

\bibitem[Houck et al.(2005)]{houck05} Houck, J.~R., et al. 2005, ApJL, in press

\bibitem[Kim \& Sanders(1998)]{kim98}
Kim,~D.-C. \& Sanders,~D.B. 1998, ApJS, 508, 627

\bibitem[Lacy et al.(2005)]{lacy05}
Lacy, M. et al. 2005, ApJ, in press

\bibitem[Laurent et al.(2000)]{laurent00} Laurent, O., Mirabel,
I.~F., Charmandaris, V., Gallais, P., Madden, S.~C., Sauvage, M., Vigroux,
L., \& Cesarsky, C.\ 2000, \aap, 359, 887

\bibitem[Lutz et al.(2005)]{lutz05} Lutz, D., Yan, L. et al. ApJ, 2005, in prep.

\bibitem[Lutz et al.(1998)]{lutz98} Lutz, D., Spoon, H.~W.~W.,
Rigopoulou, D., Moorwood, A.~F.~M., \& Genzel, R.\ 1998, \apjl, 505, L103

\bibitem[Madau et al.(1996)]{madau96} Madau, P., Ferguson, 
H.~C., Dickinson, M.~E., Giavalisco, M., Steidel, C.~C., \& Fruchter, A.\ 
1996, \mnras, 283, 1388 

\bibitem[Marleau et al.(2004)]{marleau04}
Marleau, F.~R., et al. 2004, ApJS, 154, 66

\bibitem[Papovich et al.(2004)]{casey04} Papovich, C., et al.\
2004, \apjs, 154, 70

\bibitem[Puget et al.(1996)]{puget96}
Puget,J.-L.;Abergel, A.; Bernard,J.-P. et al. 1996, A\&A, 308L, 5

\bibitem[Rigopoulou et al.(1999)]{dim99}
Rigopoulou,D.;Spoon,H.W.W.;Genzel,R.;Lutz,D.;Moorwood,A.F.M.;Tran,Q.D. 1999, AJ, 118, 2625

\bibitem[Serjeant et al.(2000)]{Serjeant00}
Serjeant et al. 2000, MNRAS, 317, 29

\bibitem[Smail, Ivison \& Blain(1997)]{smail97}
Smail, I.; Ivison,R.J.;Blain,A.W. 1997, ApJL, 490, 5

\bibitem[Soifer et al.(1987)]{tom87} Soifer, B.~T., Sanders,
D.~B., Madore, B.~F., Neugebauer, G., Danielson, G.~E., Elias, J.~H.,
Lonsdale, C.~J., \& Rice, W.~L.\ 1987, \apj, 320, 238

\bibitem[Spoon et al.(2004)]{spoon04}
Spoon, H.W.W. et al. 2004, ApJS, 154, 184

\bibitem[Spoon et al.(2004)]{spoon04b} Spoon, H.~W.~W.,
Moorwood, A.~F.~M., Lutz, D., Tielens, A.~G.~G.~M., Siebenmorgen, R., \&
Keane, J.~V.\ 2004, \aap, 414, 873

\bibitem[Tran et al.(2001)]{tran01} Tran, Q.~D., et al.\ 2001, 
\apj, 552, 527 

\bibitem[Vermeij \& van der Hulst(2002)]{vermeij02} Vermeij, R.,
\& van der Hulst, J.~M.\ 2002, \aap, 391, 1081

\bibitem[Voit(1992)]{voit92} Voit, G.~M.\ 1992, \apj, 399, 495 

\bibitem[Werner et al.(2004)]{werner04} Werner, M.~W., et al.\
2004, \apjs, 154, 1

\bibitem[Yan et al.(2004)]{yan04}
Yan, Lin et al. 2004, ApJS, 154, 60

\end{thebibliography}
\end{document}